
%
\newbox\grsign \setbox\grsign=\hbox{$>$} \newdimen\grdimen \grdimen=\ht\grsign
\newbox\simlessbox \newbox\simgreatbox
\setbox\simgreatbox=\hbox{\raise.5ex\hbox{$>$}\llap
     {\lower.5ex\hbox{$\sim$}}}\ht1=\grdimen\dp1=0pt
\setbox\simlessbox=\hbox{\raise.5ex\hbox{$<$}\llap
          {\lower.5ex\hbox{$\sim$}}}\ht2=\grdimen\dp2=0pt

\magnification=1200
\baselineskip=19pt
\vsize=8.8truein
\hsize=6.5truein
\hoffset=1.0truecm
\voffset=1.0truecm
\def\ref#1{{\par\noindent \hangindent=3em \hangafter=1 #1\par}}
\centerline{\bf THE DIFFERENCE BETWEEN RADIO-LOUD}
\centerline{\bf AND RADIO-QUIET ACTIVE GALAXIES}
\vskip 1.0cm
\centerline{A. S. WILSON}
\centerline{Space Telescope Science Institute}
\centerline{3700 San Martin Drive, Baltimore, MD 21218}
\centerline{and}
\centerline{Astronomy Department, University of Maryland,}
\centerline{College Park, MD 20742}
\vskip 0.3cm
\centerline{and}
\vskip 0.3cm
\centerline{E. J. M. COLBERT}
\centerline{Astronomy Department, University of Maryland,}
\centerline{College Park, MD 20742}
\vfill\eject

\vskip 1.0cm
\centerline {ABSTRACT}
\vskip 0.2cm
The recent development of unified theories of active galactic nuclei (AGN)
has indicated that there are two physically distinct classes of these
objects - radio-loud and radio-quiet. The primary observational
distinctions between the two types are: (1) The radio-loud objects produce
large scale radio jets and lobes, with the kinetic power of the jets being
a significant fraction of the total bolometric luminosity. On the other hand,
the weak radio ejecta of the radio-quiet objects are energetically
insignificant.
(2) The radio-loud objects are
associated with elliptical galaxies which have undergone recent mergers,
while the radio-quiets prefer spiral hosts. (3) The space density of the
radio-louds at a given optical luminosity is $\approx$ 10 times
lower than that of the radio-quiets. Despite these differences,
the (probably) thermal emissions from
the AGN (continua and lines from X-ray to infrared wavelengths) are quite
similar in the two classes of object. We argue that this last result suggests
that the black hole masses and mass accretion rates in the two classes are
not greatly different, and that the difference between the classes is
associated
with the spin of the black hole.

We assume that the normal process of accretion through a disk does not lead
to rapidly spinning holes, and propose instead that galaxies
(e.g. spirals) which have not
suffered a recent major merger event
contain non-rotating or only slowly
rotating black holes. When two such galaxies merge, the two black holes
are known to form a binary and we assume that they eventually coalesce.
In the small fraction of mergers in which the two ``parent'' galaxies
contain very massive holes of roughly equal mass, a rapidly spinning,
very massive
hole results. It is proposed that such mergers are
the progenitors of powerful radio sources, in which the radio jets are
powered by the spin energy of the
merged hole. We calculate the distributions of
mass and spin for the merged holes from the
parent hole mass distribution, which is derived from
the optical luminosity function of radio-quiet AGN adopting different activity
patterns. The ratio of the number of radio-loud to radio-quiet AGN's at a given
thermal (e.g. optical) luminosity is determined by the galaxy merger rate.
The required fraction of galaxies which merge
during the average lifetime ($\approx 10^{8}$ yrs) of a radio-loud
AGN is found to be $10^{-1}$, i.e. a merger rate of 1 in $\simeq 10^{9}$ yrs.
The Blandford-Znajek
formalism is then used to predict the radio luminosity and radio luminosity
function of the merged population. Comparisons
between the predicted and observed
radio luminosity functions constrain the efficiencies with which jet
power is extracted from
the spinning hole and radio emission is produced by the jet. The cosmological
evolution of the radio properties of the radio-loud objects is related to
the increased frequency of merger events at earlier epochs.

\vskip 0.2cm
\noindent
{\it Subject headings:} black hole physics -- galaxies: active -- galaxies:
interactions -- galaxies: jets -- galaxies: nuclei -- galaxies: quasars:
general

\vfill\eject

\centerline{1. INTRODUCTION}
\vskip 0.2cm
Over the past decade, a considerable unification in our understanding of the
phenomenology of active galactic nuclei (AGN) has been achieved. Because
radiation is not emitted isotropically by these nuclei, intrinsically similar
types of AGN may appear quite different when viewed from different directions.
Two kinds of anisotropy are recognised. One kind is emission in a narrow
cone, which is almost certainly a result of beaming by a relativistic jet. The
second kind of anisotropy involves emission in a wider cone, which is a
consequence of either
shadowing of an intrinsically isotropic, optical-ultraviolet source by a
thick (or
warped) dusty torus or an intrinsically anisotropic
optical-ultraviolet emitter.
Radio-quiet objects form one class of active galaxies, with Seyfert 1's and
radio-quiet
quasars viewed close to ``pole-on'' (so the broad line region is seen
directly) and Seyfert 2's viewed more equatorially (so the broad line region
is hidden by the dusty torus). The second class comprises low power
(Fanaroff-Riley class I, FRI) radio galaxies and BL Lac objects. A BL Lac is
seen when the nucleus is viewed within the radiation beam of a relativistic
jet which fuels the large-scale radio lobes. At larger viewing angles, the
large-scale lobes dominate the radio emission and the
object is a classical FRI radio galaxy. Lastly, OVV quasars, radio-loud
quasars,
broad-line and narrow-line radio galaxies (FRII) appear to form a third class,
with the sequence representing increasing angular difference between our line
of
sight and the axis of the source. In this class, both kinds of anisotropy
are believed to be present. These issues have been recently reviewed by
Antonucci (1993).

The reason for the large range in radio powers is unknown.
There has, however, been considerable discussion concerning whether the
radio-quiet and
radio-loud objects form distinct and separate classes, or whether there is a
continuum
of radio properties in just one class of object.
Several arguments favor the
former hypothesis, as we now summarise.
(i) There is a difference in the nature of
the host galaxies, at least at low redshift. Radio-loud
AGN are {\it never} found in spiral hosts, but almost always reside in
elliptical or
elliptical-like galaxies (a
few are associated with S0's).
Conversely, relatively few radio-quiet AGN are found
in ellipticals.
(ii) Imaging surveys have shown that the mean absolute magnitudes
of the underlying galaxies of radio galaxies and radio-loud
quasars
are the same ($<$M$_{V}>$ = --23.3;
H$_0$ = 50 km s$^{-1}$ Mpc$^{-1}$ and
q$_0$ = 0 are used throughout the paper),
consistent with the notion that they are identical objects and
that the observability of the quasar nucleus is determined by
variability or aspect dependent effects
(e.g. Malkan 1984; Smith et al. 1986; V\'eron-Cetty \&
Woltjer 1990; see Hutchings 1987 for a contradictory
result). In contrast, the host
galaxies of radio-quiet quasars are 0.6 -- 1.0 mag less luminous
(e.g. Smith et al. 1986; V\'eron-Cetty \&
Woltjer 1990; see also review by Woltjer 1990).
It has further
been argued (e.g. Smith et al. 1986; Hutchings, Janson \& Neff 1989) that
the host
galaxies of radio-loud quasars are ellipticals and those of radio-quiet
quasars are spirals, consistent with the distinction at low redshift. These
results show that the radio-loud and radio-quiet
classes inhabit different types of host galaxy.
(iii) The radio luminosity functions of Seyfert galaxies and radio galaxies
are distinct and do not join onto each other (Meurs \& Wilson 1984). Further,
radio-loud
AGN comprise an intrinsically rarer class of object than radio-quiets
(e.g.
Padovani 1993).
(iv) Radio surveys of optically selected quasars commonly
show a bimodal distribution of flux densities (e.g. Strittmatter et al. 1980;
Kellermann et al. 1989). Peacock, Miller \& Longair (1986) favored the
existence of two quasar populations with widely differing radio luminosities
and differing optical luminosity functions. This conclusion was based on their
finding that the radio properties of optically faint quasars conflicted with
a single population model (Schmidt 1970) in which the optical and radio
luminosities are correlated. In subsequent work, Miller, Peacock \& Mead (1990)
performed a radio survey of optically selected quasars covering a wide range
of optical luminosities, but only a narrow range of redshifts. These authors
found
strong evidence for the existence of two populations of quasar - one having
radio luminosities $L_{5 GHz} > 10^{26}$ $W Hz^{-1}$ and the
other having $L_{5 GHz} < 10^{25}$ $W Hz^{-1}$, with no
quasars of intermediate luminosity. They further concluded that the fraction
of quasars which are radio-loud is a function of redshift and probably
optical luminosity also, and that these effects may be understood in terms
of the mixing of two quasar populations with differing luminosity functions
and differing amounts of cosmological evolution.
(v) In both radio-quiet (de Bruyn \& Wilson 1978) and radio-loud
(Baum \& Heckman 1989; Rawlings \&
Saunders 1991) objects, there is a strong correlation
between the steep-spectrum radio and the total [OIII] $\lambda$5007
luminosities
These correlations follow roughly parallel loci in the radio --
[OIII] plane, but with the radio-loud objects a factor of $\approx$ 10$^{4}$
more luminous at 5 GHz than the radio-quiets for a given [OIII] luminosity.
Few objects fall between the two classes (Rawlings 1993), suggesting a genuine
dearth of ``radio-intermediate'' objects.

Observations indicate
that the properties of radio-loud and radio-quiet objects are similar
at infrared, optical, ultraviolet and X-ray wavelengths
(e.g. Steidel \& Sargent
1991; Sanders et al. 1989; Neugebauer et al. 1986). The shapes of the
local optical luminosity functions of radio-loud and radio-quiet quasars
are not significantly different (Padovani 1993), although the former function
is subject to
large observational uncertainties. There is now considerable evidence that most
of the
infrared - X-ray continuum is dominated by thermal emission which, in the
context of black hole accretion models, is related to the black hole
mass and the ratio of the accretion rate to the Eddington accretion rate
(i.e. the Eddington ratio). These similarities
of the thermal properties of the two classes suggest that the distributions
of black hole
masses and Eddington ratios are not widely different. It is then reasonable
to infer that the distinction between the radio-loud and radio-quiet classes
is related to the spin of the black hole.

The hypothesis (Toomre \&
Toomre 1972) that many
elliptical galaxies are formed by the merging of spirals is
supported by recent observations (Schweitzer 1982; Kormendy \&
Djorgovski 1989). Images of radio-loud AGN show clear evidence for
morphological peculiarities (e.g. tails, bridges, shells, dust lanes,
second nuclei) indicative of a galaxy interaction or merger involving
at least one disk galaxy (e.g. Heckman et al. 1986; Smith et al. 1986;
Hutchings 1987). Such evidence for interactions/mergers is found much less
frequently in radio-quiet objects (e.g. Hutchings \& Neff 1990). Thus, there
is clear evidence that the radio-loud AGN phenomenon is triggered by a
{\it recent} ($\le 10^{9}$ yrs ago) interaction or merger.

The present paper presents a model which attempts to account
for the dichotomy between the radio-loud and radio-quiet classes. We adopt
the standard supermassive black hole (SMBH) model of nuclear activity
(e.g. Rees 1984). In essence, the thermal emission
(optical-ultraviolet continuum, at
least some of the X-rays, and infrared reradiation from dust grains) is
powered by accretion.
On the other hand,
the mechanical output
of the AGN (i.e. the radio jets) is
considered to be powered by the spin of the
black hole (e.g. Blandford 1990 and references therein).
We shall, however, assume that the normal process of accretion through
an accretion
disk does {\it not} lead to a rapidly spinning SMBH, because either the
accretion
is intermittent (and residual coupling between the hole and surrounding gas
can spin down the hole between the accretion episodes) or the accretion is not
confined to a single plane
over the long timescales
(10$^{8}$ yrs for a 1 M$_\odot$ yr$^{-1}$ accretion rate onto a
10$^{8}$ M$_\odot$ hole) required to spin up the hole. Rather, {\it SMBH's
acquire a
significant fraction of the maximal Kerr angular momentum only through
the essentially instantaneous event of coalescence
with a second black hole of similar mass}.

We consider mergers in a population
of parent galaxies each of which contains a single, nuclear SMBH with small
or zero spin. In a galaxy merger, the two nuclear SMBH's form a
binary and eventually coalesce. The time
scale for this coalescence is uncertain (Begelman, Blandford \& Rees 1980),
but we shall assume it is very much less than the Hubble time.
Qualitatively, there are three possible outcomes of such merger. Two
small (say $\simeq$ 10$^{5}$ M$_\odot$) black
holes of approximately equal mass will
coalesce into a rapidly spinning, low mass hole. The merging of a large
(say $\simeq$ 10$^{8}$ M$_\odot$) and a small hole will result in a slowly
spinning, high mass hole. Lastly, the merging of two large holes will give
a rapidly spinning, high mass hole. We propose that the radio galaxy and radio
quasar phenomenon is a result of the last type of merger. The mass spectrum
of SMBH's in the parent population may reasonably be expected to decline
towards higher mass, so most mergers will involve relatively low mass holes
and will not result in a high radio luminosity object.
Because of the requirement that the holes are both massive and of comparable
mass, the powerful radio galaxy/quasar phenomenon is intrinsically rare, and
only results from a very small fraction of galaxy mergers.

In Section 2, we calculate the distributions of SMBH mass and
jet luminosity for the merged population, while Section 3 discusses the
inputs to the model in an astrophysical context. The results are compared
with observations in Section 4 and discussed in Section 5. Concluding
remarks and some outstanding issues and problems are given in Section 6.
\bigskip
\centerline{2. THE MODEL}

We consider a parent population of galaxies each of which contains a
slowly rotating or
non-rotating SMBH of mass $M$ in its nucleus. The number
density of SMBH's
in the parent population in the mass range $M$ to $M+dM$
is $N(M)dM$ Mpc$^{-3}$ in
the mass range $M_{l}$ to $M_{u}$.
A fraction $g$ of these galaxies is envisaged
to merge with another galaxy in a given time and the associated black holes
coalesce on a
timescale $\ll$ the Hubble time. We
neglect the loss of mass due to gravitational
radiation and also assume that there is no spatial mass segregation of the
parent population of SMBH's.
\smallskip
\centerline{\it 2.1 Distribution of Mass after Merging}
\smallskip
We use the subscript $m$ to represent parameters after merging.  Two parents
with masses $M_a$ and $M_b$ combine to form a daughter with mass $M_m$.
The number density of merged SMBH's in the daughter population in the mass
range $M_m$ to $M_m+dM_m$ is $N_m(M_m)dM_m$ Mpc$^{-3}$ with
$N_m(M_m)$ given by
$$\eqalign{
N_m(M_m)  &= {{g} \over {2N_o}}\int_{M_1}^{M_2} N(M)N(M_m-M)dM \cr
	  &= {{g} \over {N_o}}\int_{M_1}^{M_{m}/2} N(M)N(M_m-M)dM,\cr
} \eqno(1) $$
where $N_o = \int_{M_l}^{M_u} N(M)dM$ is~the
total number of SMBH's per Mpc$^{3}$ of the parents,
$M_{2}$ is the smaller of $M_{u}$ and $M_{m}-M_{l}$,
and $M_{1}$ is
the larger of $M_{l}$ and $M_{m}-M_{u}$.

For a power-law distribution
$N(M) = KM^{-\gamma}$ ($K$ is a normalization constant) and substituting
$x = M/M_{m}$, we have
$$\eqalign{
N_m(M_m) &= {{gK^2} \over {2N_o}} M_m^{-2\gamma+1}\int_{x_1(M_m)}^{x_2(M_m)}
       	      x^{-\gamma}(1-x)^{-\gamma}dx\cr
         &= {{gK^2} \over {N_o}} M_m^{-2\gamma+1}\int_{x_1(M_m)}^{1/2}
	      x^{-\gamma}(1-x)^{-\gamma}dx,\cr
}\eqno(2)$$
where $x_{2}$ is the smaller of $M_{u}/M_{m}$ and
$(M_{m}-M_{l})/M_{m}$, and $x_{1}$ is the larger of
$M_{l}/M_{m}$ and $(M_{m}-M_{u})/M_{m}$.
The daughter distribution in the mass range
$2M_{l} \ll M_{m} < M_{u}+M_{l}$ (i.e. $x_{1} = M_{l}/M_{m} \ll 1/2,$
and $x_{2} = 1- M_{l}/M_{m} \simeq 1$) does not show end effects of the
mass range of the parents, and in this range we have
$$N_m(M_m) = {{gK^2} \over {2N_o}} M_m^{-2\gamma+1}B(1-\gamma,1-\gamma)
\qquad {\rm for}~ \gamma < 1,\eqno(3)$$
\noindent
where $B$ is the Beta function, and
$$N_m(M_m) = g K  M_m^{-\gamma} \qquad {\rm for}~\gamma > 1.\eqno(4)$$
\noindent
Thus the slope of the daughter mass function is identical to that of the
parent if $\gamma > 1$.

The integral for $N_m(M_m)$ may alternatively be expressed in terms of the
mass ratio of the two merging black holes $r = M_{a}/M_{b}$
$$N_m(M_{m})  = {{g}M_{m} \over {N_o}}\int_{r_{1}(M_{m})}^{1}
N\biggl[{M_{m}r \over (1+r)}\biggr] N\biggl[{M_{m} \over (1+r)}\biggr]
{dr \over (1+r)^{2}}, \eqno(5)$$
where $r_{1}$ is the larger of $M_{l}/(M_{m}-M_{l})$ and
$(M_{m}-M_{u})/M_{u}$.
We have taken $r < 1$ ($M_{a} < M_{b}$) in order to avoid double-counting.
For the power-law distribution
$N(M) = KM^{-\gamma}$, equation (5) becomes
$$N_m(M_m) = {{gK^2} \over {N_o}} M_m^{-2\gamma+1}\int_{r_{1}(M_{m})}^{1}
\biggl[{r \over (1+r)^{2}}\biggr]^{-\gamma} {dr \over (1+r)^{2}}. \eqno(6)$$
Fig. 1 is a plot of lines of constant $M_{m}$ and $r$ in
the $M_{a}$--$M_{b}$ plane.  Each value of $M_m$ describes a curve for which
a specific range of $r$, corresponding to the range of integration in
equation (5), is allowed.
\smallskip
\centerline{\it 2.2 Distribution of Jet Luminosity after Merging}
\smallskip
The non-thermal emission is assumed to be powered by the spin energy
of the merged SMBH. Theories of the physical processes by which power
is extracted from the spinning hole are still in their infancy, so our
calculation of the jet power must be considered preliminary and
illustrative.
It is believed that energy may be extracted from a spinning black hole via the
magnetic field by the Blandford-Znajek process
(Blandford \& Znajek 1977). The electromagnetic power available from
this process depends on the hole's mass and spin, and the axial magnetic
field generated by external currents.
Under the assumption that the magnetic
energy density near the black hole is proportional to the shear stress
in the accretion disk,
the resulting luminosity available for extraction is conjectured
(Blandford 1990) to be
$$L_{em} = \alpha_H \lambda^2 L_{Edd}.\eqno(7)$$
where $\alpha_H$ is an unknown constant,
$L_{Edd}$ is the Eddington luminosity, and $\lambda$ is the black hole spin
parameter (see Blandford 1990 and references therein). In this formalism,
the relationship between the ``electromagnetic mass'' $M_{em}$ (defined as
$M_{em} = \lambda^2 M_{m}$)
and the jet luminosity
($L_{em} = \alpha_H {{4 \pi G m_p c} \over {\sigma_{T}}}M_{em}$)
is
analogous to that between $M_{m}$ and the Eddington
luminosity. We shall use equation (7) to obtain the jet luminosity,
while recognising the idealised nature of the expression.

We thus need to obtain the spin parameter $\lambda$ of the daughter black
hole from the merging of the two parent black holes, which are assumed to
possess no spin.
The parameter $\lambda$ is a measure of the spin per unit mass of the daughter
and so, assuming a circular orbit for the pre-merged holes, $\lambda$ is
completely determined by the ratio $r$ of the masses
of the parents.
We may write $\lambda$ in terms of the
specific angular momentum, $a$, and the gravitational radius,
$m = GM/c^{2}$, of the SMBH (cf. Blandford 1990; in relativists units
[$G = c = 1$], both $a$ and $m$ have units of length, so $a/m$ is
dimensionless):
$$ \lambda =
{  {{a}\over{m}} \over { 1 + \sqrt{1-({{a}\over{m}})^2} }  }.  \eqno(8)$$
%
The calculation of ${{a}\over{m}}$ from the coalescence of two black holes
with arbitrary mass ratio is a difficult problem in general relativity
that is currently unsolved.  However, the functional forms for $\lambda$
and ${{a}\over{m}}$ can be constrained using physical arguments.
For reasons stated in Appendix A, we have chosen the following
parameterization for ${{a}\over{m}}(r)$:
$${{a}\over{m}}(r) \approx
\biggl(1 - {{4f_1r}\over{(1+r)^2}}\biggr)
\biggl[2\sqrt{3} + (3.95 - 2\sqrt{3})
{{4r}\over{(1+r)^2}}\biggr]
{{r}\over{(1+r)^2}}, \eqno(9)$$
where $f_1$ is the fraction of angular momentum at the last stable
circular orbit which is lost to gravitational radiation during the
plunge to coalescence for equal-mass parents.
For small values of $r$, $\lambda(r)$ has the limiting form $\sqrt{3}r$.

Sample plots of $\lambda(r)$ are shown in Fig. 2.
We have experimented with other reasonable (see
Appendix A) estimates of the form for $\lambda(r)$ when
calculating the distribution of jet luminosities (see below), and note
there are no significant effects on our results.
We vary $f_1$ from 0.0 to 0.5
in the following calculations.

The distribution of  $M_{em}$ after merging
can be written
$$N_m(M_{em}) = {{g} \over {N_o}}
\int_{r_{1}(\lambda, M_{em})}^{r_{2}(\lambda, M_{em})}
{N[{r M_{em} \over {\lambda^{2}(r) (1+r)}}]}
{N[{M_{em} \over {\lambda^{2}(r) (1+r)}}]} {M_{em}
\over {\lambda^{4}(r) (1+r)^{2}}} dr.\eqno(10)$$
The lower limit, $r_{1}$, is determined by solving
$\lambda^{2}(r_{1}) (1+r_{1}) = M_{em}/M_{u}$. The upper limit, $r_{2}$,
is either 1 or the solution of
$\lambda^{2}(r_{2}) (1+r_{2})/r_{2} = M_{em}/M_{l}$, whichever is smaller.
The changeover between these two values for $r_{2}$ occurs at
$M_{em, crit} = 2\lambda^{2}(1)M_{l}$.
Fig. 3
shows lines of constant
$M_{em}$ and $r$ in the $M_{a}$ vs $M_{b}$ plane. As in Fig. 1, each value
of $M_{em}$ describes a curve for which a specific range of $r$,
corresponding to the range of integration in equation (10), is
allowed.

For a power-law parent mass function $N(M) = KM^{-\gamma}$, equation (10)
becomes
$$N_m(M_{em}) = {{gK^2} \over {N_o}} M_{em}^{-2\gamma+1}
\int_{r_{1}(\lambda, M_{em})}^{r_{2}(\lambda, M_{em})}
r^{-\gamma} \lambda^{4\gamma-4}(r) (1+r)^{2\gamma-2} dr.\eqno(11)$$
\noindent
The integration in equation (11) can be performed analytically in
limiting cases, and results are summarized in Appendix B.
\bigskip
\centerline{3. ESTIMATES OF IMPORTANT PARAMETERS}
\vskip 0.2cm
In order to apply the model described in Section 2 and derive the properties
of the merged (``daughter'') population, we need to estimate a number of
required parameters. We first describe how the parent mass function is
obtained.
Then, in order to guide the comparison
of the model with observations (Section 4),
we also estimate canonical
values for $g$, $\alpha_{H}$, and $\epsilon$ (the efficiency of conversion of
jet power to radio emission).
\vskip 0.2cm
\centerline{\it 3.1 The Parent Mass Distribution}
\vskip 0.2cm
This distribution may be estimated from the optical luminosity function (OLF)
of AGN (e.g. Marshall 1985, 1986; Boyle, Shanks \& Peterson 1988) by
making assumptions about the bolometric correction (i.e. the ratio of total
thermal luminosity to that in the B band), the Eddington ratio and the
activity patterns of the AGN's over cosmological timescales. In
particular, Cavaliere \& Padovani (1988, 1989) have derived mass distributions
under three assumptions concerning the activity patterns. At one extreme is a
pattern (which they label `C'
for `continuous') of literal luminosity evolution, with each
object dimming continuously over the timescale ($\approx$ 3 Gyr) for
evolution of the quasar population. High redshift quasars evolve to local
Seyfert 1's in a one-to-one relationship, and AGN activity is hosted by a
privileged set of some 1\% of bright galaxies. A second, intermediate
pattern (called `R' for `recurrent' by Cavaliere \& Padovani)
is provided by discontinuous
activity in a larger fraction of galaxies, consisting of short, recurrent
episodes of activity. This behavior could be driven by, for example,
episodes of accretion stimulated by interaction of the host galaxy with
companion. In the third pattern (labelled `E' for
`Eddington'), which is at the opposite
extreme to
`C', activity represents a single, short event in the lifetime of every
galaxy.

In pattern `C', $L/L_{E}$ is estimated to be
$\simeq 2\times10^{-4} \eta_{E,-1}$ at the
present epoch, where $\eta_{E,-1}$ is the efficiency of
mass - energy conversion, in units of $10^{-1}$. A small fraction of
galaxies (i.e. those which exhibit activity) contain very massive
($10^{10} - 10^{12}$ M$_\odot$) remnants. For the case of pattern `R',
Cavaliere \& Padovani (1988) favor
$L/L_{E}$ $\simeq 5\times10^{-2} \eta_{E,-1}$.
The black hole masses for both currently active and inactive galaxies cover
a range $\approx 10^{7} - 10^{9}$ M$_\odot$, with currently inactive galaxies
having a space density $\simeq$ 500 times that of currently active ones
(see Fig. 1 of Cavaliere \& Padovani 1988).
For pattern `E' in its simplest
form, L/L$_{E}$ is taken to be $\simeq$ 1 and all galaxies contain SMBH with
masses ranging between $\simeq 10^{6}$ and $10^{10}$ M$_\odot$. To illustrate
the application of our model, we shall use (Section 4) the Cavaliere \&
Padovani mass functions for cases `C' and `R', derived using
L/L$_{E}$ $\simeq$ 10$^{-4}$ and L/L$_{E}$ $\simeq$ 5$\times$10$^{-2}$,
respectively.
\vskip 0.2cm
\centerline{\it 3.2 The Merger Rate}
\vskip 0.2cm
The parameter $g$ represents the fraction of SMBH's in a unit volume of space
that merge in the relevant time interval. Because the merged population
of interest comprises those galaxies currently active as radio-loud AGN's,
this time interval is the average lifetime of the extended radio sources,
which is commonly estimated to be $\tau_{R} \simeq 10^{8}$ years.
The fraction of black holes which merge in this time depends on both the
galaxy merger rate and the time required for the two SMBH's to coalesce.
Both of these timescales are controversial. The merger rate is undoubtedly
a function of both the environment and the mass ratio of the two galaxies.
Schweizer (1982, 1990) argues that all ellipticals may have undergone at least
one merger per Hubble time $\tau_{H}$. Merger rates of 1 per
$10^{9}$ yrs may apply to close binary galaxies (Borne 1984) or groups
(Navarro, Mosconi \& Garcia Lambas 1987). Roos (1981) estimates a merger rate
of 1 per $10^{9}$ yrs when the second galaxy has 1\% of the mass of the first.
Heckman, Carty \& Bothun (1985) have shown that the average local galaxy
density is larger (by at least a factor of 2 -- 3) for radio-loud ellipticals
and lenticular galaxies than for radio-quiet galaxies of the same morphological
type. Therefore, it is likely that the somewhat higher merger rates
(i.e. 1 in $10^{9}$ yrs) apply to radio galaxies.
The time to
coalescence, $\tau_{C}$, of the SMBH's has been
variously claimed to be as short as $\approx 10^{7}$ years (Ebisuzaki,
Makino \& Okumura 1991) or longer than the Hubble time
(e.g. Valtaoja, Valtonen \& Byrd 1989). In our model, we obviously
require $\tau_{C}$ $<<$ $\tau_{H}$, and probably
$\tau_{C}$ $\le$ $10^{9}$ yrs (so that the radio galaxy is
recognisable as a {\it recent} merger).
Writing $\tau_{M}$ = $10^{9}$ yrs for the average time
between mergers, we take as
our canonical value $g$ $\simeq$ $\tau_{R}$/$\tau_{M}$
$\simeq$ $10^{8}$/$10^{9}$ $\simeq$ $0.1$.

For $\tau_{R} \simeq 0.1\tau_{M}$, it is unlikely that the merged galaxy
will suffer a second merger during its lifetime as a radio source.
Therefore, we do not need to consider the
effect of a {\it second} black hole merger event on the mass
and spin of the merged hole during the lifetime of most
radio sources.

\vskip 0.2cm
\centerline{\it 3.3 The Electromagnetic Extraction Efficiency}
\vskip 0.2cm
The parameter $\alpha_{H}$ is a measure of the efficiency of
conversion of the spin energy of the daughter SMBH to
jet power. We shall adopt
$\alpha_{H}$ = 1 as a canonical value.
\vskip 0.2cm
\centerline{\it 3.4 The Efficiency of Conversion of Jet Power to Radio
Emission}
\vskip 0.2cm
Here the typically quoted numbers of
$\epsilon = L_{radio}/L_{em} \simeq  10^{-1} - 10^{-2}$
may be considered canonical (see Leahy 1991, who also discusses the
limitations of this
form of parameterisation of the radio source energetics).

The effects of changes in these parameters on the properties of the OLF
and radio luminosity function
(RLF) of the merged (radio-loud) population are
easily gauged. Increases in N(M) (i.e. a greater density of parent holes
of a given mass) and $g$ increase the daughter number density at a given
mass and luminosity. Increasing M increases the optical and radio
luminosities, while altering $\alpha_{H}$ and $\epsilon$ affects only the
radio luminosities. In fact, the radio luminosities depend on the
{\it product} $\alpha_{H}\epsilon$, not on $\alpha_{H}$ and $\epsilon$
separately.
\bigskip
\centerline{4. COMPARISON WITH OBSERVATIONS}
\vskip 0.2cm
In comparing our model with the OLF and RLF of
active galaxies, we have proceeded as follows.
\smallskip
\noindent
(a) A `parent' population is first adopted. In the case of `C' type activity,
only the currently active galaxies are considered to contain SMBH's, so the
appropriate parent is
the radio-quiet population. Because the space density of
radio-louds is much less than that of radio-quiets, we can, to a good
approximation, use the total (radio-quiet plus radio-loud) population as the
parent. For `R' type activity, the {\it whole} galaxy
population, which is dominated by currently inactive galaxies (see Fig. 1 of
Cavaliere \& Padovani 1988), should be taken as the parent.
\smallskip
\noindent
(b) To derive the parent mass distribution, we first fitted the B band
OLF N$_{tot}(M_{B}$)
of the total population of AGN (Padovani 1993;
Cheng et al. 1985) with a two power law
form (cf. Marshall 1986; Boyle, Shanks \& Peterson 1988). The function
chosen was:
$$
N(L_B) = {{N^*}\over{L_B^*}} \times \left\{\eqalign{
{{L_B} \overwithdelims () {L_B^*}}^{-2.2} \qquad L &< L_B^* \cr
{{L_B} \overwithdelims () {L_B^*}}^{-3.5} \qquad L &\ge L_B^* \cr
} \right],
\eqno(12)
$$
where $N^* =$ 5 $\times$ 10$^{-8}$ Mpc$^{-3}$.  The luminosity function
changes slope at $L_B^* =$ 5.8 $\times$ 10$^{10}$ L$_\odot$, which
corresponds to $M_B^* =$ --23.65.
The data and
the fitted function are plotted in Fig. 4. By adopting a value, b, for the
fraction of thermal emission which is radiated in the B band
($b \simeq 0.03$; Blandford 1990), the total thermal luminosity
corresponding to a given $M_{B}$ may be found. Given the assumed Eddington
ratios ($L/L_{E}$) adopted by Cavaliere \& Padovani (1988)
and summarised above, the distribution of black hole masses in currently active
galaxies is then calculated. For activity pattern `C', this distribution
represents the total mass distribution of SMBH's in galaxies, and is
thus assumed
to represent $N(M)$. For activity
pattern `R', we must multiply the mass function by the ratio of the total
number density of SMBH's (assumed to be dominated by currently inactive
galaxies) to the number density of currently active SMBH's.
We have used the ratio $\simeq$ 500 advocated by Cavaliere \& Padovani
(1988, 1989) to obtain $N(M)$ for pattern `R'. An important issue is the
relationship between the activity patterns of the radio-quiet and radio-loud
populations. For `C' type activity, we assume
that a radio source turns on shortly after the SMBH's have merged
and continues as a radio-loud AGN for a single
period of order 10$^{8}$ yrs (cf. Section 3.2). For `R' type activity,
the radio-quiet and radio-loud patterns are assumed to
be similar: only
1 in 500 of the merged holes is radio active at any given time and, when
active, the lifetime of the activity is the canonical 10$^{8}$ yrs.
Physically, the inactive times would correspond to periods without gaseous
accretion; such accretion is necessary both for fuelling the thermal emission
and for providing the magnetic field used to tap the spin energy of
the hole.
\smallskip
\noindent
(c) The distribution of merged mass, $N_{m}(M_{m})$, was then
calculated using equation (5). In fact, the ratio $N_{m}(M_{m})/N(M) = g$
when $M_{m} = M$ (equation 4). If we assume that the
values of $b$ and $L/L_{E}$ are the same for the radio-loud and radio-quiet
objects, then $N_{rl}(M_{B})/N_{rq}(M_{B}) = g$, independent of $b$ and
$L/L_{E}$. Here $N_{rl}(M_{B})$
is the OLF of the radio-louds and
$N_{rq}(M_{B})$ that of the radio-quiets.
By fitting our calculated $N_{rl}(M_{B})$
to Padovani's measurements, we obtain $g = 0.1$ (see Fig. 4), although
the slightly higher value of $g = 0.3$ may apply at high optical luminosities
($M_{B} < -24$). For a radio galaxy lifetime of $\simeq 10^{8}$ yrs, the
implied rate of merging is 1 in $10^{9}$ yrs, in excellent agreement
with that obtained from other considerations in Section 3.2.
\smallskip
\noindent
(d) Given the parent mass distribution $N(M)$ (see (b) above), the value of
$g$ just obtained, and the assumed forms of $\lambda(r)$
(equations 8 and 9), we
calculate $N_{m}(M_{em})$ using equation (10). $N(L_{em})$ is then
found (using
$L_{em} = \alpha_H {{4 \pi G m_p c} \over {\sigma_{T}}}M_{em}$; we
have taken $\alpha_H = 1$), with
$L_{em}$ assumed to represent the jet luminosity. Adopting a value
for the efficiency, $\epsilon$, of conversion of jet power to radio
emission and a form for the synchrotron spectrum (a power law
$S \propto \nu^{-\alpha}$ with $\alpha = 0.75$ between 10 MHz and
100 GHz), we finally obtain the RLF
of radio-loud AGN in terms of $P_{20}$, the monochromatic radio power
at 20 cm.
The predicted curves are plotted along with the
observationally determined RLF's of radio galaxies and radio-loud
quasars in Fig. 5.
\bigskip
\centerline{5. DISCUSSION}
\vskip 0.2cm
We have found that our model can account for the present day relative numbers
of radio-loud and radio-quiet AGN's if the merger rate is 1 in 10$^{9}$ yrs.
However, if such a high rate were sustained over a Hubble time and is
applicable everywhere, the parents
(spirals) would
have long since disappeared.
It is known that the galaxies hosting radio-quiet
AGN's tend to be in low galaxy density environments whereas those hosting
radio-loud objects are found preferentially in high density environments
(e.g. Smith \& Heckman 1990). A more sophisticated calculation would allow
for a range of surrounding galaxy densities and hence merger rates; radio-quiet
(unmerged) objects would naturally tend to be found in low density
and radio-loud (merged) in high density regions. Elliptical-spiral mergers
are likely to be important in the denser regions.

The good fit to the RLF shown in Fig. 5 is obtained
through mergers involving the low mass (M$_{B} >$ --24) portion of the parent
mass function (Fig. 4). In this region, the mass function of the parents
is $N(M) \propto M^{-2.2}$, which implies a RLF
$N_{m}(P_{20}) \propto P_{20}^{-3.4}$ (equation B4), in good agreement
with the RLF observed above the break near
$log P_{20} = 25$ $W Hz^{-1}$. Mergers involving the high mass part of
the parent mass function, which follows $N(M) \propto M^{-3.5}$, are
predicted (equation B4) to give $N_{m}(P_{20}) \propto P_{20}^{-6}$, a much
steeper form than observed in any part of the RLF. This steep predicted form at
high masses is intrinsic to the conjecture (equation 7) used to
calculate the radio luminosities. The formal implication is that black hole
masses in FRII radio galaxies should be similar to those in radio-quiet
objects with $M_{B} > -24$, rather than those with $M_{B} < -24$.

If
the parent mass distribution, $N(M)$, has a sharp cut-off at masses below
those probed by the OLF (Fig. 4),
the predicted RLF function for
$log P_{20}$ $<$ $10^{26}$ $W Hz^{-1}$ (corresponding to
$M_{em}$ $<$ $M_{em, crit} = 2\lambda^{2}(1)M_{l}$) has the form
$N_{m}(P_{20}) \propto P_{20}^{0.2}$ (equation B5). This last form does not
fit the data
near or below the break in the RLF, suggesting
there is a low luminosity turnover, rather than a sharp
cut-off, in the OLF at $M_{B} > -18$. Alternatively, it would be possible
to make an {\it ad hoc} modification of the form of equation (7) to allow
the break in the OLF (i.e. the parent mass function)
to correspond to the break in the RLF.

In the context of the current model, cosmological evolution of the radio
luminosities would most naturally be ascribed to a higher
black hole merger rate at earlier epochs. Such would provide density evolution,
but radio luminosity evolution could also occur if $\alpha_{H}$ and/or
$\epsilon$ are epoch dependent.
\bigskip
\centerline{6. CONCLUDING REMARKS AND FUTURE WORK}
\vskip 0.2cm
In this paper, we have assumed that radio-loud AGN have rapidly spinning,
supermassive black holes (SMBH's) in their nuclei and that the radio jets are
powered by the spin energy of the holes. Contrary to the conventional view
that SMBH's are spun up via gaseous accretion through a disk, we argue that
such holes acquire a large spin only through a merger with another SMBH.
Presumably low mass SMBH's are more common in galactic nuclei than high mass
ones, so most mergers will involve relatively low mass SMBH's. Let us also
suppose that the mass of the parent galaxy is correlated with the mass
of its nuclear
SMBH. Bearing in mind that only mergers between galaxies
of approximately equal mass are likely to give elliptical-like remnants, we
may qualitatively distinguish three types of mergers, depending on the
masses of the two SMBH's (and hence parent galaxies) involved.
\vskip 0.1cm
\noindent
(i) Low mass - low mass coalescences will lead to a rapidly spinning, low mass
hole in an elliptical-like galaxy. Such objects are probably those ellipticals
with low luminosity, pc-scale radio cores, which are common in early-type
galaxies (e.g. Sadler et al. 1994).
\vskip 0.1cm
\noindent
(ii) Low mass - high mass coalescences result in a slowly spinning, high mass
hole. The merged galaxy will retain the morphological type of the more massive
galaxy and such galaxies will form part of the radio-quiet population.
However, these objects are expected to comprise only a small fraction of the
radio-quiet population, which is dominated by the unmerged galaxies
(parents).
\vskip 0.1cm
\noindent
(iii) High mass - high mass coalescences form a rapidly spinning, high mass
hole. We associate these rare events with the formation of a powerful radio
source.

Our idealised calculations show that both the optical and radio luminosity
functions of powerful radio galaxies can be reproduced by a simple model.
Merger rates of 1 per 10$^{9}$ yrs are required in the regions where FRII
radio galaxies form. The product of the parameters $\alpha_{H}$
(defined by $\alpha_{H}$ = $L_{em}/ \lambda^2(r) L_{Edd}$, where
$L_{em}$ is the jet luminosity, $\lambda$ the black hole spin
parameter, and $L_{Edd}$ is the Eddington luminosity) and
$\epsilon$ ($\epsilon$ = $L_{radio}/L_{em}$) is constrained to
be $\approx$ 5$\times10^{-2}$ -- $10^{-4}$, depending on which pattern
of evolution of AGN activity is chosen (Section 4).

In summary, our model accounts naturally for the following observations:
\vskip 0.1cm
\noindent
(1) No spiral galaxies host radio-loud AGN.
\vskip 0.1cm
\noindent
(2) Powerful radio galaxies and quasars are the products of a recent
merger.
\vskip 0.1cm
\noindent
(3) Only a tiny fraction of galaxy mergers produce a radio galaxy (despite
the belief that SMBH's are common in galactic nuclei).
\vskip 0.1cm
\noindent
(4) Radio-loud AGN are much rarer than radio-quiet AGN.

We conclude by listing a number of issues and problems which require
further work.
\vskip 0.1cm
\noindent
(a) Measurements of the galaxy density and merger rate
in the environment of powerful radio
sources are needed. Further observational work on FRII
radio galaxies is particularly desirable. It is, of course, possible that
some radio galaxies are the result of the merger of an isolated binary galaxy.
In that case, the density of galaxies around the radio galaxy after
the merger would be
a poor indicator of the relevant merger rate.
\vskip 0.1cm
\noindent
(b) Further calculations of the timescale for coalescence
of two nuclear SMBH's in a galaxy merger due
to stellar
and gas dynamical processes are important.
\vskip 0.1cm
\noindent
(c) The controversial (e.g. Punsly
\& Coroniti 1990) topic of energy extraction from spinning SMBH's deserves
more study,
in order to
provide a firmer estimate of the jet power.

We thank R. D. Blandford, J. P. Henry,
A. Illarionov, T. Jacobson, L. Kidder, M. Livio, E. Poisson,
and B. C. Whitmore for
valuable discussions. This work was supported in part by NASA grants
NAGW-2689 and NAGW-3268.
\vfill\eject

\centerline{REFERENCES}
\vskip 0.5cm

\ref{Antonucci, R. R. J. 1993, ARAA, 31, 473}

\ref{Auriemma, C., Perola, G. C., Ekers, R., Fanti, R., Lari, C.,
Jaffe, W. J. \& Ulrich, M. H. 1977, A\&A, 57, 41}

\ref{Baum, S. A. \& Heckman, T. M. 1989, ApJ, 336, 702}

\ref{Begelman, M. C., Blandford, R. D. \& Rees, M. J. 1980, Nature, 287, 307}

\ref{Blandford, R. D. 1990, In Active Galactic Nuclei, ed T. J.-L. Courvoisier
\& M. Mayor, Saas-Fee Advanced Course 20 (Berlin: Springer-Verlag), 161}

\ref{Blandford, R. D. \& Znajek, R. L. 1977, MNRAS, 179, 433}

\ref{Borne, K. D. 1984, ApJ, 287, 503}

\ref{Boyle, B. J., Shanks, T. \& Peterson, B. A. 1988, MNRAS, 235, 935}

\ref{de Bruyn, A. G. \& Wilson, A. S. 1978, A\&A, 64, 433}

\ref{Cavaliere, A. \& Padovani, P. 1988, ApJ, 333, L33}

\ref{Cavaliere, A. \& Padovani, P. 1989, ApJ, 340, L5}

\ref{Cheng, F., Danese, L., de Zotti, G. \& Franceschini, A. 1985,
MNRAS, 212, 857}

\ref{Ebisuzaki, T., Makino, J. \& Okumura, S. K. 1991, Nature, 354, 212}

\ref{Edelson, R. A. \& Malkan, M. A. 1986, ApJ, 308, 59}

\ref{Elvis, M., Green, R. F., Bechtold, J., Schmidt, M., Neugebauer, G.
Soifer, B. T., Matthews, K. \& Fabbiano, G. 1986, ApJ, 310, 291}

\ref{Fanti, R. \& Perola, G. C. 1976, In Radio Astronomy \& Cosmology,
IAU Symposium Nr. 74, ed D. L. Jauncey p171 (Dordrecht: Reidel)}


\ref{Heckman, T. M., Carty, T. J. \& Bothun, G. D. 1985, ApJ, 288, 122}

\ref{Heckman, T. M. et al. 1986, ApJ, 311, 526}

\ref{Hutchings, J. B. 1987, ApJ, 320, 122}

\ref{Hutchings, J. B., Janson, T. \& Neff, S. G. 1989, ApJ, 342, 660}

\ref{Hutchings, J. B. \& Neff, S. G. 1990, AJ, 99, 1715}

\ref{Kellermann, K. I., Sramek, R. A., Schmidt, M., Shaffer, D. B., \&
Green, R. 1989, AJ, 98, 1195}

\ref{Kidder, L. E., Will, C. M., \& Wiseman, A. G. 1993,
Physical Review D, 47, 8, 3281}

\ref{Kormendy, J. \& Djorgovski, S. 1989, ARAA, 27, 235}

\ref{Leahy, J. P. 1991, In Beams and Jets in Astrophysics, Ed. Philip A.
Hughes, p 100 (Cambridge University Press)}

\ref{Malkan, M. A. 1984, ApJ, 287, 555}

\ref{Marshall, H. L. 1985, ApJ, 299, 109}

\ref{Marshall, H. L. 1986, In Structure and Evolution of Active Galactic
Nuclei,
Eds G. Giuricin et al., p 627 (Reidel: Dordrecht)}

\ref{Meurs, E. J. A., \& Wilson, A. S. 1984, A\&A, 136, 206}

\ref{Miller, L., Peacock, J. A., \& Mead, A. R. G. 1990, MNRAS, 244, 207}

\ref{Navarro, J. F., Mosconi, M. B., \& Garcia Lambas, D. 1987,
MNRAS, 228, 501}

\ref{Neugebauer, G., Miley, G. K. Soifer, B. T. \& Clegg, P. E. 1986,
ApJ, 308, 815}

\ref{Padovani, P. 1993, MNRAS, 263, 461}

\ref{Peacock, J. A., Miller, L., \& Longair, M. S. 1986, MNRAS, 218, 265}

\ref{Punsly, B. \& Coroniti, F. V. 1990, ApJ, 350, 518}

\ref{Rawlings, S. \& Saunders, R. 1991, Nature, 349, 138}

\ref{Rawlings, S. 1994, In The First Stromlo Symposium: The Physics of Active
Galaxies, ASP Conference Series Vol. 54, Eds G. V. Bicknell, M. A. Dopita,
\& P. J. Quinn, p 253 (Astronomical Society of the Pacific: San Francisco)}

\ref{Rees, M. J. 1984, ARAA, 22, 471}

\ref{Roos, N. 1981, A\&A 104, 218}

\ref{Sadler, E. M., Slee, O. B., Reynolds, J. E. \& Ekers, R. D. 1994,
In The First Stromlo Symposium: The Physics of Active
Galaxies, ASP Conference Series Vol. 54, Eds G. V. Bicknell, M. A. Dopita,
\& P. J. Quinn, p 335 (Astronomical Society of the Pacific: San Francisco)}

\ref{Sanders, D. B., Phinney, E. S., Neugebauer, G., Soifer, B. T. \&
Matthews, K. 1989, ApJ, 347, 29}

\ref{Schmidt, M. 1970, ApJ, 162, 371}

\ref{Schweizer, F. 1982, ApJ, 252, 455}

\ref{Schweizer, F. 1990, Talk given at STScI Workshop,
``The Galaxy Merger Rate''}

\ref{Smith, E. P., Heckman, T. M., Bothun, G. D., Romanishin, W. R.,
\& Balick, B. 1986, ApJ, 306, 64}

\ref{Smith, E. P. \& Heckman, T. M. 1990, ApJ, 348, 38}

\ref{Steidel, C. C. \& Sargent, W. L. W. 1991, ApJ, 382, 433}

\ref{Strittmatter, P. A., Hill, P., Pauliny-Toth, I. I. K., Steppe, H.,
\& Witzel, A. 1980, A\&A, 88, L12}

\ref{Toomre, A., \& Toomre, J. 1972, ApJ, 178, 623}

\ref{Valtaoja, L., Valtonen, M. J., \& Byrd, G. G. 1989, ApJ, 343, 47}

\ref{V\'eron-Cetty, M. P. \& Woltjer, L. 1990, A\&A, 236, 69}

\ref{Wall, J. V., Pearson, T. J. \& Longair, M. S. 1980, MNRAS, 193, 683}

\ref{Woltjer, L. 1990, In Active Galactic Nuclei, ed T. J.-L. Courvoisier
\& M. Mayor, Saas-Fee Advanced Course 20, p1 (Berlin: Springer-Verlag)}

\vfill\eject

\centerline{\bf Figure Captions}
\vskip 0.5cm

\noindent
Figure 1.  Contours of the merged mass $M_m$ (solid) and the mass ratio
$r$ (dashed) in the $M_a$--$M_b$ plane for the case
$M_l$ = $10^{5}$ M$_\odot$, $M_u$ = $10^{9}$ M$_\odot$.  In order to avoid
double-counting, we have chosen the upper limit of integration in $r$ ($r < 1$)
from the condition $M_a < M_b$.  The dotted contour is $M_m$ = $M_u + M_l$.
When $M_m$ $<$ $M_u + M_l$,
the lower limit of integration in equations (5) and (6) is determined from
$M_a = M_l$,
i.e. $r_{1}$ = $M_l$/($M_m - M_l$).
When $M_m$ $>$ $M_u + M_l$, the lower limit is determined from
$M_b = M_u$, i.e. $r_{1}$ =  ($M_m - M_u$)/$M_u$.

\noindent
\bigskip

\noindent
Figure 2. The spin parameter $\lambda$ as a function
of the mass ratio $r$ for $f_1 =$ 0.0 -- 0.5.  The limiting form for small
$r$ ($\lambda(r) = \sqrt{3}r$) is shown as a dashed line.

\bigskip

\noindent
Figure 3. Contours of the merged electromagnetic mass $M_{em}$
(solid and dotted) and the
mass ratio $r$ (dashed) in the $M_a$--$M_b$ plane.
The solid lines represent constant $M_{em}$ for f$_{1}$ = 0,
while the dotted represent constant $M_{em}$ for f$_{1}$ = 0.5.
The lower limit of
integration ($r_{1}$) in equations (10) and (11)
is determined from $M_b = M_u$.  The upper limit ($r_{2}$) is
determined by $M_a = M_b$ ($r = 1$) for
$M_{em}$ $>$ $M_{em, crit} = 2\lambda^{2}(1)M_{l}$, and by
$M_a = M_{l}$ for $M_{em}$ $<$ $M_{em, crit}$.

\bigskip

\noindent
Figure 4. Optical Luminosity Functions of Active Galaxies.  Filled squares
represent all quasars in the Padovani (1993)
sample, while filled triangles correspond to Seyfert galaxy nuclei
(Cheng et al. 1985).
Open squares correspond to the radio-loud quasars in Padovani's sample.
The parent mass function was computed by fitting the
luminosity function
for the whole sample (upper line) with a two power-law form (see
Section 4).
The merger rate of $g = 10^{-1}$
was chosen so that the
predicted luminosity function for the daughter (merged)
population (lower curve) agrees with the data for the radio-loud quasars
(Section 4).

\bigskip

\noindent
Figure 5.  Radio Luminosity Functions of Active Galaxies
for $f_1 =$ 0.0 -- 0.5. The symbols
represent the observationally inferred radio luminosity functions
for radio galaxies (Auriemma et al. 1977 -- squares), radio galaxies
plus radio-loud quasars (Wall, Pearson
\& Longair 1980 -- triangles) and radio-loud quasars (Fanti \& Perola
1976 -- circles).
The dashed lines represent extensions of the parent
mass function beyond the range in $M_B$ of the
Cheng et al. (1985) and Padovani (1993) samples (see
Figure 4). For the `C' type models, the plotted curves correspond to
$\epsilon$ = 10$^{-4}$, while for the `R' type,
$\epsilon$ = 5$\times$10$^{-2}$.

\bigskip

\noindent
Figure A1. The spin ${{a}\over{m}}$ as a function of the mass ratio $r$ for
$f_1 =$ 0.0 -- 0.5. The limiting form for small $r$ is shown as a
dashed line.

\bigskip

\noindent
Figure A2. The derivative of the spin parameter $\lambda$ with respect
to the mass ratio $r$ as a function of $r$ for $f_1 =$ 0.0 -- 0.5.

\vfill\eject
\centerline{\bf Appendix A: The Spin Parameter $\lambda$}
\vskip 0.2cm
The spin parameter $\lambda$ and ${{a}\over{m}}$ are related through
the transformation
$$ \lambda = {  {{a}\over{m}} \over { 1 + \sqrt{1-({{a}\over{m}})^2} }  }
\quad \quad \quad
{{a}\over{m}} = {{2\lambda}\over{1+\lambda^2}}.  \eqno(A1)$$
The problem of finding the resulting spin ${{a}\over{m}}$ as a function
of $r$, the ratio of the masses of the two parent black holes,
involves general relativistic computer codes and a solution is not yet
available.
A recent study by Kidder, Will \& Wiseman (1993) evolves the
binary system down to the innermost stable circular orbit (ISCO) for
arbitrary mass ratio.
They find (see their Fig. 3) that the angular momentum per unit reduced
mass ($\tilde{J}$)
at ISCO is approximately
linear in the parameter $\eta$ (the ratio of the reduced mass to the total
mass), rising from ${{\tilde{J}}\over{m}} = 2\sqrt{3}$ for the test-mass
case ($r$$\to$$0$, $\eta$$\to$$0$) to
${{\tilde{J}}\over{m}} = 3.95$ in the case of equal masses
($r = 1$, $\eta = 0.25$).
The parameter $\eta$ can be written in terms of the mass ratio as
$\eta = r/(1+r)^2$.
In order to bridge the gap between ${{\tilde{J}}\over{m}}$ at ISCO and
${{a}\over{m}}$ after coalescence, we
introduce a function $f$ (the fraction of angular momentum
at ISCO that is lost in gravitational radiation
during the plunge from ISCO to coalescence), which is also a function
of the mass ratio $r$.  Then we can write
$$
{{a}\over{m}}(r) = (1-f(r)) {{\tilde{J}}\overwithdelims (){m}}_{ISCO} \eta
\eqno(A2)$$
$$~~~~~~\approx (1-f(r)) \biggl[2\sqrt{3} + (3.95 - 2\sqrt{3})
{{4r}\over{(1+r)^2}}\biggr] {{r}\over{(1+r)^2}}.
$$

We may place restrictions on $f$, ${{a}\over{m}}$ or $\lambda$ in order to
constrain the functional form $\lambda(r)$.  We must have
$$0 < {{a}\over{m}}(r) < 1 \quad {\rm and} \quad 0 < \lambda(r) < 1
\quad{\rm for}\quad 0 < r < 1. \eqno(i)$$

It is obvious that we must also have
$$0 < f(r) < 1 \quad{\rm for}\quad 0 < r < 1. \eqno(ii)$$
With this restriction, the requirement that ${{a}\over{m}} < 1$
always holds true.

In the test-mass limit,
we assume that the loss of angular momentum during the
plunge is negligible ($f(0) = 0$), which yields a limiting form
for $\lambda(r)$:
$$\lambda(r) =  {{1}\over{2}} {{a}\over{m}}(r) = \sqrt{3}r
\quad\quad {\rm as} \quad r \to 0.  \eqno(iii)$$

One expects the spin to increase monotonically with the ratio of the masses
of the parent black holes, which implies an additional requirement of
$$
{{d {{a}\over{m}} }\over{dr}} > 0, \quad {\rm or,~equivalently,} \quad
{{d\lambda}\over{dr}} > 0 \quad  \quad 0 < r < 1. \eqno(iv)$$

Various functional forms for $f(r)$ were tried.  A simple linear form
$f(r) = a + br$ ($a =$0, $b <$ 1) fails criterion $(iv)$.  Other forms
({\it e.g.} polynomial, polynomial/exponential combination) can be
constructed to satisfy
$(i-iv)$, but such forms require either a large number of
parameters (in which case interpreting variations in a single parameter is
not straightforward) or a complex functional form.
A linear function in $\eta$ ($f(r) = a + b\eta$; $a =$ 0, $b <$ 4)
does satisfy $(i-iv)$ if $b < $2.1, so we have
used the single-parameter form
$$ f(r) = {{4f_1r}\over{(1+r)^2}} \eqno(A3)$$
in our calculations (section 2.2).
Here $f_1 = b/4 = f(1)$ is the fraction of angular momentum at ISCO
lost to gravitational radiation during the plunge in an equal-mass system.
The resulting spin ${{a}\over{m}}$ and spin parameter $\lambda$ are
shown in Figs. A1 and 2, respectively, for values of $f_1 =$ 0.0 -- 0.5.
For all values of $f_1$, ${{d\lambda}\over{dr}} = \sqrt{3}$ at $r = 0$
and ${{d\lambda}\over{dr}} > 0$ for all $r$ (see Fig. A2).

\vfill\eject
\centerline{\bf Appendix B: Analytic Forms for the Distribution of
$M_{em}$}
\vskip 0.2cm
Equation (11) can be simplified in limiting cases. In the following, we have
assumed $\lambda(r) = \sqrt{3}r$ for all $r$:
\smallskip
\noindent
$\gamma < 1$ and $3M_{l}^{2}/M_{u} \ll M_{em} \ll M_{u}:$
$$N_m(M_{em}) = {{gK^{2}3^{2\gamma-3}} \over {(1-\gamma)N_o}}
({{1} \over {3M_{u}}})^{{3\gamma-3} \over {2}}
M_{em}^{{-(\gamma+1)} \over {2}}. \eqno(B1)$$
\noindent
$\gamma = 1$ and $M_{em, crit} < M_{em} \ll M_{u}:$
$$N_m(M_{em}) = {{gK^{2}3^{2\gamma-2}} \over {N_o}} {M_{em}}^{-1}
\ln({\sqrt{{3M_{u}} \over {M_{em}}}}). \eqno(B2)$$
\noindent
$\gamma = 1$ and $3M_{l}^{2}/M_{u} \ll M_{em} < M_{em, crit}:$
$$N_m(M_{em}) = {{gK^{2}3^{2\gamma-2}} \over {N_o}} {M_{em}}^{-1}
\ln({{1} \over {M_{l}}} \sqrt{{M_{em}M_{u}} \over {3}}). \eqno(B3)$$
\noindent
$\gamma > 1$ and $M_{em, crit} < M_{em} \ll M_{u}:$
$$N_m(M_{em}) = {{gK^{2}3^{2\gamma-3}} \over {(\gamma-1)N_o}}
{M_{em}}^{-2\gamma+1}. \eqno(B4)$$
\noindent
$\gamma > 1$ and $3M_{l}^{2}/M_{u} \ll M_{em} < M_{em, crit}:$
$$N_m(M_{em}) = {{gK^{2}3^{2\gamma-3}} \over {(\gamma-1)N_o}}
({{1} \over {3M_{l}}})^{3\gamma-3} M_{em}^{\gamma-2}. \eqno(B5)$$

\vfill\eject
\end